\documentclass[aps,pra,longbibliography,twocolumn]{revtex4-2}
\usepackage{amsmath}
\usepackage{amssymb}
\usepackage{amsthm}
\usepackage{setspace}
\usepackage{graphicx}
\usepackage{braket}
\usepackage{mathrsfs}
\usepackage{float}
\usepackage[dvipsnames]{xcolor}
\definecolor{electricpurple}{rgb}{0.75, 0.0, 1.0}
\usepackage[colorlinks = true, linkcolor=blue,citecolor=magenta,urlcolor=electricpurple,anchorcolor = blue]{hyperref}
\usepackage[utf8]{inputenc}
\usepackage[english]{babel}
\usepackage{bm}
\usepackage[normalem]{ulem}

\newcommand{\ttd}[1]{}
\newcommand{\ttds}[1]{{\color[rgb]{0,0,0}{#1}}}

\newcommand{\beq}{\begin{equation}}
\newcommand{\eeq}{\end{equation}}
\newcommand{\bea}{\begin{eqnarray}}
\newcommand{\eea}{\end{eqnarray}}

\newcommand{\bk} { \bm{k} }

\DeclareMathOperator{\sech}{sech}
\newcommand{\eqn}[1] {Eq.~(\ref{#1})}
\newcommand{\fig}[1]{Fig.~\ref{#1}}
\newcommand{\mylabel}[1]{\label{#1}}

\begin{document}

%\preprint{AIP/123-QED}

\title{Signature of point nodal superconductivity in the Dirac semimetal PdTe}
% Force line breaks with \\
\author{C. S. Yadav$^{1}$}
\email{shekhar@iitmandi.ac.in}
\author{Sudeep Kumar Ghosh$^{2}$}
\email{skghosh@iitk.ac.in}
\author{Pankaj Kumar$^{1}$}
\author{A. Thamizhavel$^{3}$}
\author{P.L. Paulose$^{3}$}

\affiliation{$^{1}$School of Physical Sciences, Indian Institute of Technology Mandi, Kamand, Mandi-175075 (H.P.) India}%

\affiliation{$^{2}$Department of Physics, Indian Institute of Technology Kanpur, Kanpur- 208016 (U.P.), India }

\affiliation{$^{3}$DCMP \& MS, Tata Institute of Fundamental Research, Mumbai -400005 (Maharashtra) India}

\date{\today}

\begin{abstract}

Recent Angle-Resolved Photo-emission Spectroscopy (ARPES) experiments [Phys. Rev. Lett. 130, 046402 (2023)] on PdTe, a 3D-Dirac semimetal and a superconductor with the transition temperature T$_{\rm c} \sim$ 4.3 K, have revealed compelling evidence of the presence of bulk nodes in the superconducting order parameter. To investigate the validity of this proposition, here we present a detailed investigation of the magnetic field dependence of the specific heat of PdTe down to temperatures $\sim$ 58 mK. We observed that the low temperature specific heat of PdTe with an externally applied magnetic field exhibits a power-law field dependence, a characteristic of unconventional superconductivity. Furthermore, the zero-field low-temperature electronic specific heat follows a cubic temperature dependence, which is a signature of the presence of bulk point nodes in PdTe. These intriguing observations suggest that PdTe is a rare and fascinating topological material that exhibits both Dirac semimetallic properties and superconductivity with point nodal gap symmetry.

\end{abstract}

\maketitle

\section{Introduction}

Topological semimetals represent a fascinating class of materials characterized by topologically nontrivial bulk band crossings and robust surface states that exhibit unique electronic properties, making them highly desirable for practical applications~\cite{Lv2021}. The realm of topological semimetals becomes even more captivating when combined with superconductivity, as it opens up the possibility of realizing intrinsic topological superconductivity. Recent discoveries of materials exhibiting both topological semimetallic behaviour and superconductivity have further bolstered this research direction, providing new avenues for exploring and understanding the intriguing interplay between topology and superconductivity. Examples include, superconducting 3D Dirac semimetals (Ta,Nb)OsSi~\cite{Ghosh2022}, 3D Weyl semimetals LaPt(Si,Ge) ~\cite{Shang2022}, $\mathbb{Z}_2$ TMs (Cs,Rb,K)V$_3$Sb$_5$~\cite{Jiang2022,Shan2022} and ZrOsSi~\cite{Ghosh2023}, Weyl nodal-line SMs La(Ni,Pt)Si~\cite{shang2022spin} and Kramers nodal-line SMs (Ti, Nb, Hf, Ta)RuSi~\cite{Shang2022}.

Among the transition metal chalcogenides, FeSe shows unconventional superconductivity below $\sim 8.2$ K. The Scanning Tunneling Microscopy (STM) experiments on FeSe crystalline films suggested the existence of nodal superconductivity in stoichiometric FeSe~\cite{song2011direct}. Furthermore, the identification of C$_4$ rotation-symmetry protected Dirac semimetallic behavior, featuring distinctive surface Dirac cones, in the Fe(Se,Te) superconductors~\cite{zhang2018observation, zhang2019multiple} has significantly advanced research in this class of materials. Several theoretical investigations~\cite{kawakami2019topological,wu2022nodal,kheirkhah2022surface} on these compounds have emphasized the potential for realizing various topological phases in Dirac semimetal superconductors. These phases include topological crystalline superconductivity~\cite{kawakami2019topological}, nodal higher-order topological superconductivity~\cite{wu2022nodal} and, novel gapped superconducting state with surface Bogoliubov-Dirac cones and a second-order time-reversal invariant topological superconductor with helical Majorana hinge modes~\cite{kheirkhah2022surface}. Recent angle-resolved photo-emission spectroscopy (ARPES) measurements~\cite{yang2023coexistence} provide compelling evidence that the chalcogenide superconductor PdTe exemplifies such intriguing material characteristics {as well}.

PdTe is a hexagonal superconductor with T$_{\rm c} \sim 4.3$ K \cite{karki2012pdte, karki2013interplay, tiwari2015pdte} that has a layered structure similar to that of the tetragonal Fe chalcogenide superconductors. However, the $4d$-orbital character of Pd and the face-shared octahedral environment suppress electronic correlations and promote three-dimensionality in PdTe~\cite{ekuma2013first,chapai2023evidence}. This enhanced three-dimensionality, in turn, gives rise to weaker Fermi surface nesting and diminished long-range correlations~\cite{ekuma2013first}. Based on first principles calculations, it was proposed that there is a diminished charge transfer energy between Pd-$d$ and Te-$p$ states in PdTe when compared to Fe chalcogenides. The Pd-$d$ states in PdTe are almost fully occupied, leaving no states available to form a large Hund's moment and there is no long-ranged magnetic correlations present in PdTe~\cite{ekuma2013first}. 

%%%%%%%%%%%%%%%%%%%%%%%%%%%%%%%%%%%%%%%%%%%%%%%%%%%%%%%%%%%%%%%%%%%%%%%
%%%%%%%%%%%%%%%%%%%%%%%%%%%%%%%%%%%%%%%%%%%%%%%%%%%%%%%%%%%%%%%%%%%%%%%
\begin{figure*}[htb]
	\includegraphics[width= 2.1\columnwidth]{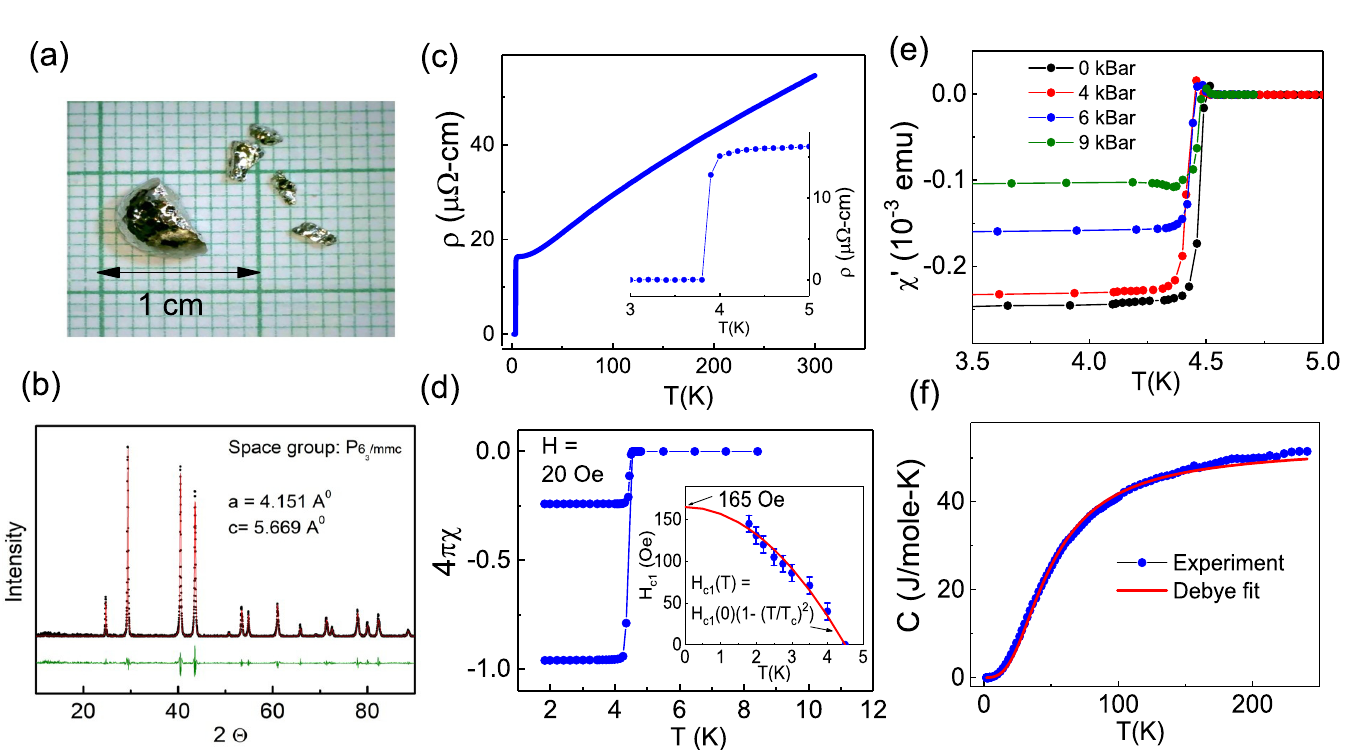}
	\caption{\textbf{Characterization of PdTe samples}: (a) An image of the PdTe ingot showing the shiny nature of crystallites. (b) XRD pattern of powdered PdTe sample showing a clean hexagonal phase (space group: P$6_{3}{/mmc}$). (c) The temperature dependence of the electrical resistivity. The inset shows a zoomed region around the superconducting transition (1.8 $\leq$ T $\leq$ 5 K). (d) DC magnetic susceptibility as a function of temperature at H = 20 Oe. The inset shows $H_{c1}(T)$ obtained from the $M-H$ isotherms, and the fit with $H_{c1}(T) = H_{c1}(0)(1-(T/T_C)^2)$. (e) AC magnetic susceptibility measured at different external hydrostatic pressures \textit{p} = 0, 4, 6, and 9 kbar. (f) Total specific heat C(T) fitted with the Debye model (\eqn{eqn:eq1}).}
	\label{fig:characterization}
\end{figure*} 
%%%%%%%%%%%%%%%%%%%%%%%%%%%%%%%%%%%%%%%%%%%%%%%%%%%%%%%%%%%%%%%%%%%%%%%
%%%%%%%%%%%%%%%%%%%%%%%%%%%%%%%%%%%%%%%%%%%%%%%%%%%%%%%%%%%%%%%%%%%%%%%

In recent ARPES measurements on PdTe~\cite{yang2023coexistence}, a noteworthy revelation emerged that a bulk Dirac point is situated just below the Fermi level accompanied by Fermi arc surface states on the $(010)$ surface. These investigations also uncovered compelling evidence of a rare phenomenon: the coexistence of nodeless superconductivity on the surface and nodal superconductivity in the bulk of PdTe. Inspired by the distinctive characteristics exhibited by this material, here we embarked on comprehensive field- and temperature-dependent investigations of the heat capacity of PdTe, down to ultra-low temperatures ($\sim 58$mK). We found that the zero-field electronic specific heat data at these low temperatures in the superconducting state aligns remarkably well with the Bardeen-Cooper-Schrieffer (BCS)-type model featuring a p-wave gap symmetry with point nodes in the bulk. Furthermore, our findings gain additional support from the magnetic field dependence of the electronic heat capacity, which distinctly points toward the unconventional nature of superconductivity in PdTe.

\section{Results and Discussion}

PdTe samples were prepared by solid-state reaction method using high purity ($\geq 99.9 \%$) Pd and Te powders inside an evacuated quartz tubes at 1080$^{0}$C for 24 hours. The ingot were found to consist of large crystallites of 1-2 mm size as shown in \fig{fig:characterization}(a). The Rietveld refinement of the X-ray diffraction (XRD) data shown in \fig{fig:characterization}b revealed the formation of a clean NiAs-type hexagonal phase in PdTe with the space group symmetry P$6_{3}{/mmc}$ (No. 194) and with the lattice parameters, a = 4.15 $\textup{\AA}$ and c = 5.67 $\textup{\AA}$.

The electrical characterizations (under an ac current 10 mA at frequency 79 Hz) and the heat capacity measurements were performed using a Physical Properties measurement System (PPMS), and the magnetic measurements were performed using a Magnetic property measurement system (SQUID-MPMS). The temperature dependence of the electrical resistivity $\rho(T)$ of PdTe is shown in \fig{fig:characterization}(c). A sharp superconducting transition is observed at T $\sim  4.3$ K (inset of \fig{fig:characterization}(c)), with full diamagnetic signal (4$\pi\chi$ = -1) as shown in \fig{fig:characterization}(d). We note that $\rho(T)$ follows normal Fermi liquid behaviour ($\rho(T)$ $\propto$ T$^{2}$) up to temperatures $\sim 42$ K, which is more than the 1/5$^{th}$ of the Debye-temperature ($\Theta_{D}$ $\sim$ 200 K) of PdTe. The inset of \fig{fig:characterization}(d) shows temperature dependence of the lower critical field ($H_{c1} (T)$) obtained from the $M-H$ isotherms. Fitting of $H_{c1}(T)$ curve with the empirical relation  $H_{c1}(T) = H_{c1}(0)(1-(T/T_C)^2)$ gives $H_{c1}(0) \approx 165 \pm 10$ Oe, which is slightly lower than the reported value ($\approx$ 200 Oe)~\cite{tiwari2015pdte}.

The AC susceptibility data measured at different externally applied hydrostatic pressures ($\textit{p}$) = 0, 4, 6 and 8 kbar are shown in \fig{fig:characterization}(e). We note that the T$_{\rm c}$ of PdTe gets suppressed with increasing pressure ($\Delta{\rm T}_{\rm c}$/$\Delta\textit{p} \sim - 0.01$ K/kbar). The low value of the coefficient of the suppression of T$_{\rm c}$ of PdTe with $\textit{p}$ may imply that its density of states (DOS) is nearly flat at the Fermi energy. This is consistent with the band structure calculations showing that the Fermi energy is located very close to the local maxima of the DOS~\cite{ekuma2013first}. The decrease in T$_{\rm c}$ with pressure for PdTe contrasts with unusually large $\textit{p}$-dependence on T$_{\rm c}$ shown by the Fe chalcogenide superconductors~\cite{Johnston2010puzzle}. \ttds{Note that the high-pressure AC susceptibility measurements were conducted using a pressure cell, making it challenging to calculate the sample's exact demagnetization factor. While we couldn't estimate the precise superconducting volume fraction under pressure, the data clearly indicates that it decreases as pressure increases.} 

The variation of the total heat capacity (C) in zero magnetic field with temperature for PdTe is shown in shown in the \fig{fig:characterization}(f). We note that in the normal state at high temperatures the specific saturates to C(T = 250 K) $\sim$ 50 J/mole-K which is close to classical Dulong-Petit limit of $C = 3nR = 6R = 49.9$ J/mole-K due to the lattice vibrations, as expected. The low-T normal state heat capacity data is fitted with a combination of the electronic and lattice contributions (Debye model of specific heat) using the expression \cite{Kittel2004}
\begin{equation}
C = \gamma T + 9nR\left(\frac{T}{\Theta_D}\right)^3\int_0^y\frac{x^4 e^x}{(e^x-1)^2}dx
\mylabel{eqn:eq1}
\end{equation}
where \textit{y} = $\Theta_{D}$⁄T and $\Theta_{D}$ is the Debye Temperature. The best possible fit gives the Sommerfeld coefficient $\gamma \sim 6.32$ mJ/mole-K$^{2}$ and $\Theta_{D} \sim 200$ K. The values of $\gamma$ in normal state, obtained from the low Temperature linear fit of C/T vs T$^{2}$ data is also $\sim 6.30$ mJ/mole-K$^{2}$. The DOS at Fermi energy N(E$_{F}$) obtained using the relation $\gamma$ = $\pi^2$  k$_{B}^{2}$ N(E$_{F}$)/{3} is N(E$_{F}$)$\sim 2.67$ states/eV-f.u. This DOS N(E$_{F}$) contains the enhancement factor due to electron-phonon coupling constant $\lambda_{el-ph}$, and can be related to the bare-band DOS at the Fermi level N$_{0}$(E$_{F}$) by the expression  N(E$_{F}$) = N$_{0}$(E$_{F}$)(1 + $\lambda_{el-ph}$). The calculated value of the $\lambda_{\rm el-ph}$ using the MacMillan’s theory, corresponding to $\Theta_{D}$ = 200 K is $\lambda_{\rm el-ph} \sim 0.7$, which is similar to that of a weak coupling BCS superconductor~\cite{mcmillan1968transition}. However, considering the modified Macmillan formula with the logarithmic averaged phonon frequency it was reported that $\lambda_{el-ph}$ $\approx$ 1.4 in the Ref.~\cite{karki2012pdte}.

Further, we estimate the Fermi velocity ${v_F}$ in PdTe from the N(E$_{F}$) using a single band free electron gas model as $v_{F} = \frac{\pi^2\hbar^3}{m^{*2}V_{\mathrm{f.u.}}}N(E_F)$, where $V_{f.u.}$ = V$_{cell}$/2 \cite{Kittel2004}. Assuming m$^{*}$ = m$_{e}$, we have $v{_F} = 4.8 \times 10^{8}$ cm/sec. The mean free path \textit{l} = $v{_F}$ $\tau$, ($\tau$ = mean relaxation time) obtained using the formula $l=3\pi^2\left(\frac{\hbar}{e^2\rho_0}\right)\left(\frac{\hbar}{m^{*}v_F}\right)^2$ is estimated to be \textit{l} $\sim$ 43.8 nm \cite{Kittel2004}. This is an order of magnitude smaller than \textit{l} = 305.4 nm (estimated using \textit{l} = (2m$_{e} v{_F}$)/(ne$^{2}\rho_{0}$), where $n$ is carrier concentration obtained from Hall measurement and $\rho_{0}$ is the resistivity at T = 2 K~\cite{karki2012pdte}) corresponding to the $v{_F}$ $\sim$ 7 $\times$ 10$^{7}$ cm/sec estimated by considering the bare band density of states obtained from the band structure calculation reported in the Ref.~\cite{karki2012pdte}.

%%%%%%%%%%%%%%%%%%%%%%%%%%%%%%%%%%%%%%%%%%%%%%%%%%%%%%%%%%%%%%%%%%%%%%%
%%%%%%%%%%%%%%%%%%%%%%%%%%%%%%%%%%%%%%%%%%%%%%%%%%%%%%%%%%%%%%%%%%%%%%%
\begin{figure}[!b]
	\includegraphics[width=\columnwidth]{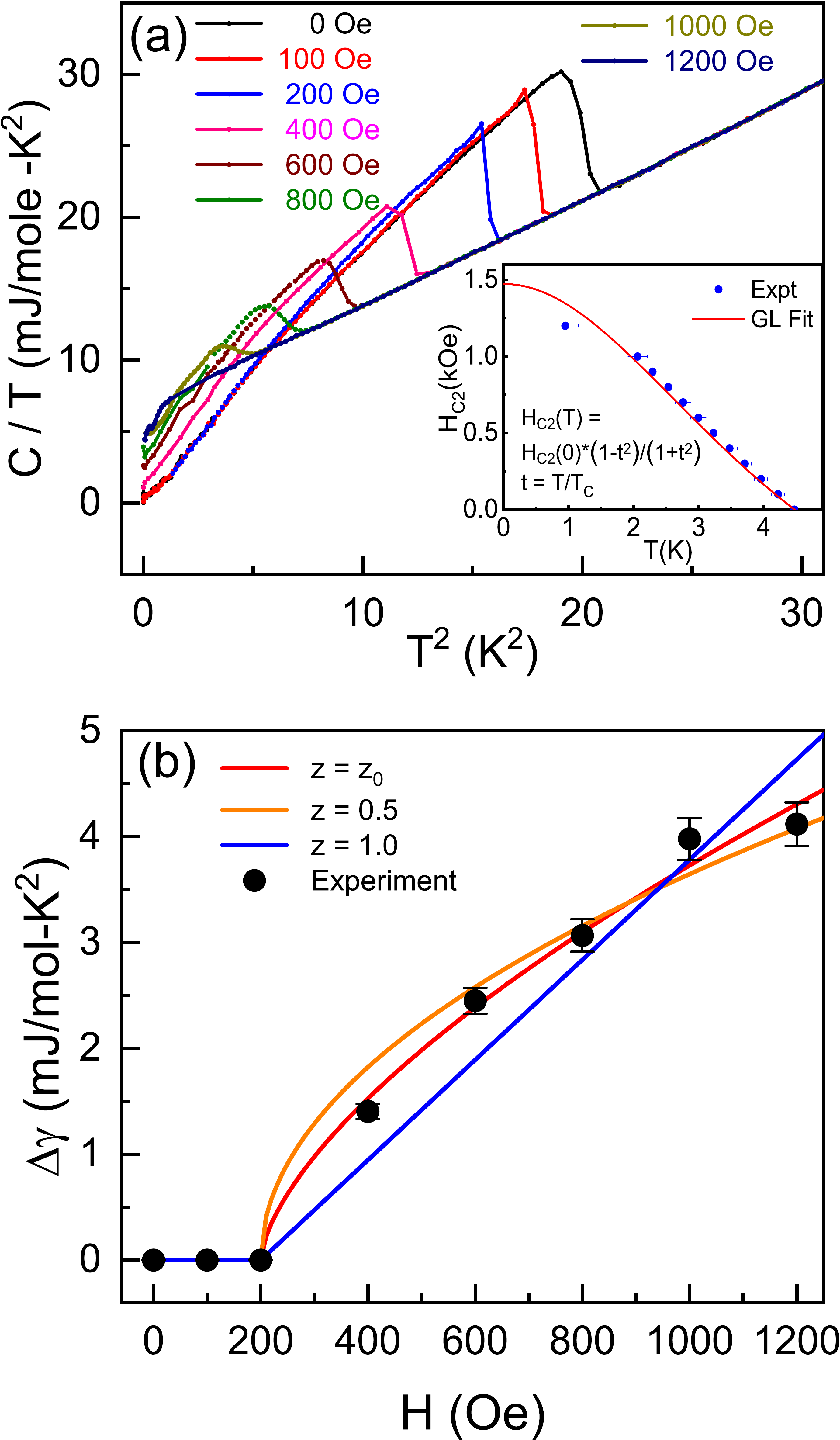}
%	\vspace{-10 pts}	
	\caption{\textbf{Field dependence of specific heat}: (a) Variation of the total specific heat (C) with temperature at various magnetic fields: H = 0 - 1200 Oe, The inset shows the temperature of the upper critical field H$_{c2}$. The dashed red line in the inset shows a fitting of the H$_{c2}$ data using the {generalized Ginzburg-Landau equation: H$_{c2}$(T) = H$_{c2}$(0)[(1-t$^{2}$)/(1+t$^{2}$)]~\cite{Maki1966,Werthamer1966,Fang2005}}. (b) Magnetic field dependence of the relative electronic specific heat $\Delta \gamma$ given by \eqn{eqn:delgamma}. The fittings shown \ttds{are with the form $\Delta \gamma \propto (H-H_0)^z$} in the mixed state of PdTe for $H > H_0$ with $H_0 \approx H_{c1}(0)$. \ttds{The best is fit found with the exponent $z = z_0 = 0.644 \pm 0.059$ which is clearly different from the linear ($z=1$) and square-root ($z=0.5$) dependences shown.}}
	\label{fig:Hcfig}
\end{figure}
%%%%%%%%%%%%%%%%%%%%%%%%%%%%%%%%%%%%%%%%%%%%%%%%%%%%%%%%%%%%%%%%%%%%%%%
%%%%%%%%%%%%%%%%%%%%%%%%%%%%%%%%%%%%%%%%%%%%%%%%%%%%%%%%%%%%%%%%%%%%%%%

Now, we focus on the low temperature ($T \sim 0.058 - 6$ K) behavior of the specific heat C(T) in externally applied magnetic fields. The variation of the specific heat in this regime as a function of temperature for different magnetic fields up to $\sim 1200$ Oe are shown in the \fig{fig:Hcfig}(a). The gradual suppression of the peak height and of T$_{\rm c}$ in the presence of magnetic field is clearly evident from this figure. The variation of the upper critical field H$_{c2}$(T) with temperature is shown in the inset of \fig{fig:Hcfig}(a). {We have fitted this H$_{c2}$(T) vs T data using the generalized Ginzburg-Landau (GL) equation: H$_{c2}$(T) = H$_{c2}$(0)[(1-t$^{2}$)/(1+t$^{2}$)]~\cite{Maki1966,Werthamer1966,Fang2005} where t = T/T$_{\rm c}$ is the reduced temperature, the zero-temperature upper critical field is H$_{c2}$(0), and T$_{\rm c}$ is 4.47 K.\ttd{\sout{This expression fits better than the usual H$_{c2}$(T) = H$_{c2}$(0) [1-(T/T$_{\rm c}$)$^{2}$] behavior expected for the conventional superconductors {near T$_{\rm c}$.}}} {The value of H$_{c2}$(0) as obtained from the fit is H$_{c2}$(0) = 1473.16 Oe. Further, we have extracted the zero-temperature orbital limited upper critical field H$_{c2}^{orb}$(0) from the Werthamer-Helfand-Hohenberg formula given as $\mu_0 H_{c2}(0) = - 0.693 T_{\rm c} (\mu_0 dH_{c2}/dT)_{T = T_{\rm c}}$ ~\cite{momono1996evidence}. The calculated value of $\mu_0 H_{c2}(0)$ is 1332 Oe, which is lower than 1473.155 Oe, obtained from the GL fit.} The coherence length $\xi$ calculated using $\xi$ = ($\phi_{0}$⁄(2$\pi$ $\mu_0$H$_{c2}$))$^{1/2} \sim 47.2$ nm which is close to the mean free path $ l \sim 43.8$ nm of PdTe indicating that PdTe is close to the dirty limit \cite{yadav2009upper, Alex2011PRB}.

We have extracted the relative electronic specific heat defined as: 
\beq \mylabel{eqn:delgamma}
\Delta\gamma (H) = \lim_{T \to 0}\left[\frac{C(H,T)}{T} - \frac{C(H=0,T)}{T} \right]
\eeq
from the low temperature specific heat data (shown in \fig{fig:Hcfig}(a)) at different magnetic fields. The field dependence of $\Delta\gamma (H)$ for type-II superconductors in the mixed state is expected to be $\propto H$ for a conventional \textit{s}-wave BCS gap and a power law $\propto H^z$ with $0.5 < z < 1$ for an unconventional gap symmetry in general~\cite{Volovik1993,Nakai2004,Bang2010,Zehetmayer2013} \ttds{with $z=0.5$ usually implying the presence of line nodes in the order parameter for a two dimensional material~\cite{Volovik1993,Nakai2004}}. The variation of $\Delta\gamma (H)$ with the applied magnetic fields in the mixed state of the PdTe superconductor is shown in \fig{fig:Hcfig}(b). We have fitted the data for $\Delta\gamma(H)$ in the mixed state for $200 \lesssim H \lesssim 1200$ Oe by a power law $\sim (H-H_0)^z$ for $H>H_0$ where $H_0 \approx H_{c1} (0)$. We find a \ttd{\sout{reasonably }}good fit with the exponent \ttds{$z = 0.644 \pm 0.059$} \ttd{\sout{close to $0.5$ }}as shown in \fig{fig:Hcfig}(b). Such a power law field dependence of $\Delta\gamma(H)$ clearly points to the unconventional nature of superconductivity in PdTe consistent with previously reported results~\cite{tiwari2015pdte}. {However, to clearly determine the nature and position of the nodes in the superconducting gap, orientation-dependent measurements of the low temperature specific heat using good quality single crystals are needed~\cite{Mirano2003, Vekhter1999}.}

The low temperature total specific heat data \ttds{$C_{\rm total}(T)$ in the zero field} of PdTe in the normal state ($T_{\rm c} \leq T \leq 10$K) is fitted with the form \ttds{$C_n(T) = \gamma_{N} T + C_{\rm ph}$}, where the first term is the electronic contribution to the specific heat \ttds{in the normal state} and the second term is the  phononic contribution to the specific heat C$_{\rm ph} = \beta$T$^{3}$ + $\delta$T$^{5}$. The fitting \ttds{shown in the inset of \fig{fig:specific_heat}(a)} gives $\gamma_{N}$ = 6.303 mJ/mole-K$^{2}$, $\beta$ = 0.748 mJ/mole-K$^{4}$ and $\delta$ = 0.0019 mJ/mole-K$^{6}$ for PdTe. Then we extract the low temperature electronic specific heat (C$_{\rm el}$) data of PdTe by defining $C_{\rm el}(T) =  C_{\rm total}(T) - C_{\rm ph}(T)$. The variation of the C$_{\rm el}$(T) at zero-field ($H = 0$) at low temperatures is shown in the \fig{fig:specific_heat}. The low temperature behaviour of C$_{\rm el}$ can be used to investigate symmetry of the superconducting gap realized in a particular material. To this end, we have defined $\gamma_e (T) ={\rm  C}_{\rm el}/T$ and fitted it  with a generalized single-band BCS-type form applicable for different types of superconducting gap symmetries given by~\cite{xing2016nodal}:
\beq
\mylabel{eqn:cv_theory}
\!\!\!\! \gamma_e \!=\! \frac{N(E_F)}{2 k_B T^3} \!\! \left\langle \! \int_0^\infty \!\!\!\! d\epsilon \sech^2\!\left(\!\frac{E_{\bk}}{k_B T}\!\right) \!\! \left[ E^2_{\bk} - \frac{T}{2}\frac{d\Delta^2_{\bk}(T)}{dT}\right] \! \right\rangle_{\!\!\!FS}\!\!
\eeq
where $E_{\bk} = \sqrt{\epsilon^{2} + \Delta^2_{\bk}(T)}$ and $\langle\rangle_{FS}$ represents an average over the Fermi surface (assumed to be spherical). The gap function ($\Delta_{\bk}(T)$) defined as $\Delta_{\bk}(T) = \Delta'(T) g(\bk)$ has a universal temperature dependence $\Delta'(T) = \Delta_0 \tanh\left[1.82 \left\{ 1.018 \left({\rm T}_{\rm c}/T -1\right)^{0.51}\right\} \right]$ and all the information about its symmetry is contained in the function $g(\bk) = g(\theta, \phi)$ with $\theta$ and $\phi$ being azimuthal and polar angles respectively~\cite{Biswas2021}. Then for a particular gap symmetry we would have only one fitting parameter $\frac{\Delta_0}{k_B {\rm T}_{\rm c}}$ in \eqn{eqn:cv_theory}. In an s-wave superconductor, the isotropic gap at the Fermi surface leads to the exponential decay of the electronic specific heat on decreasing temperature. However, this behaviour of the electronic specific heat gets affected by the presence of nodes in the superconducting gap at the Fermi surface and  for $T < {\rm T}_{\rm c}$ a power-law decay of electronic specific heat is observed that goes as $\propto T^2$ for line nodes and $\propto T^3$ in the presence of point nodes~\cite{goll2006unconventional,pines1995dx2,Ghosh2021}.

%%%%%%%%%%%%%%%%%%%%%%%%%%%%%%%%%%%%%%%%%%%%%%%%%%%%%%%%%%%%%%%%%%%%%%%
%%%%%%%%%%%%%%%%%%%%%%%%%%%%%%%%%%%%%%%%%%%%%%%%%%%%%%%%%%%%%%%%%%%%%%%
\begin{figure}[!tb]
	\includegraphics[width=\columnwidth]{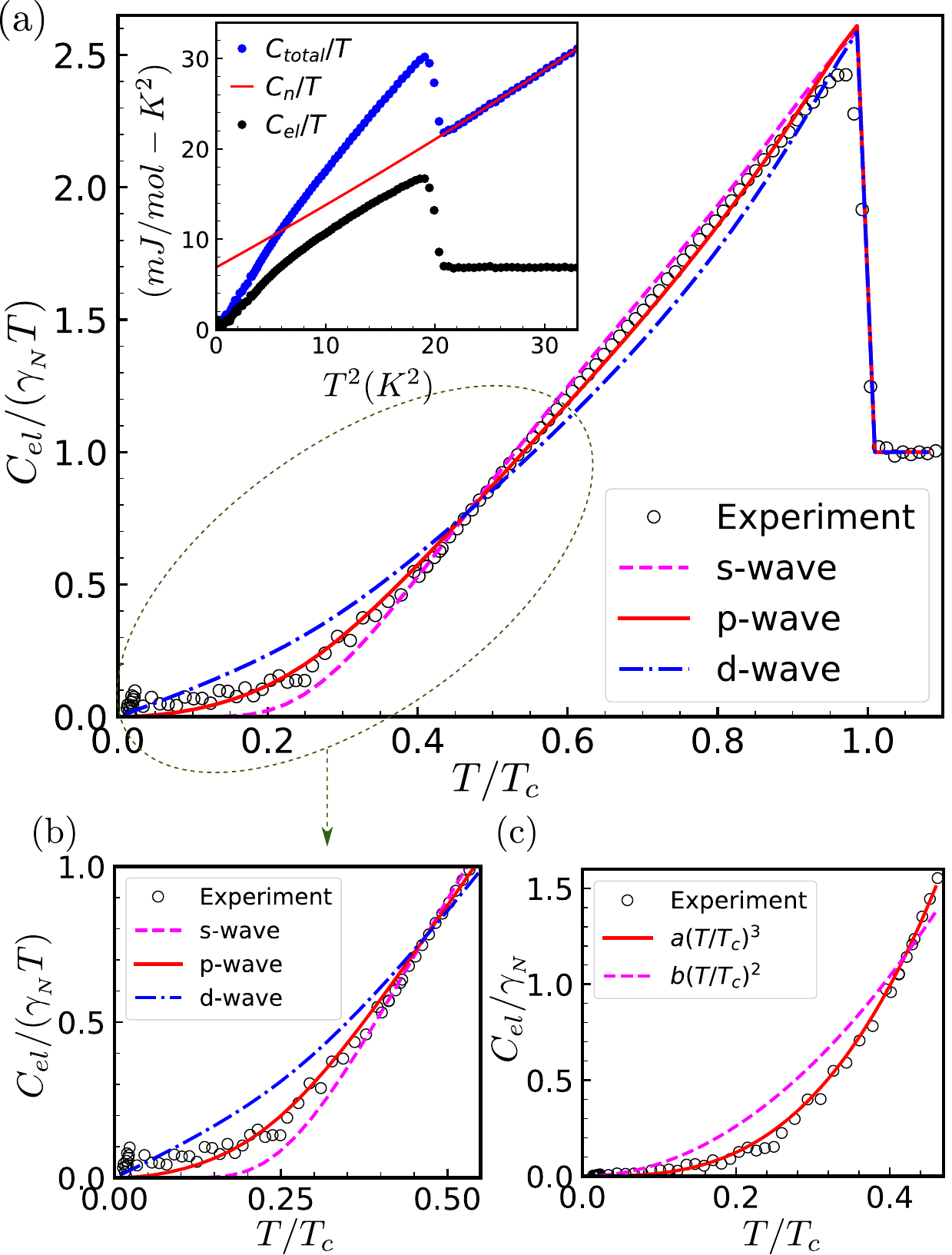}
	\caption{\textbf{Zero-field {electronic} specific heat}: (a) The electronic specific heat (C$_{el}$) data as a function of temperature without any external field present. The experimental data is fitted with {three different models of the order parameter:} a fully gapped $s$-wave BCS (pink), a point nodal $p$-wave (red) and a line-node dominated $d$-wave (blue). \ttds{The inset shows behavior of the total specific heat ($C_{total}$), the total specific heat in the normal state ($C_n$) extrapolated to zero temperature and the electronic specific heat ($C_{el}$) obtained after subtracting the phonon contribution from $C_{total}$}. {(b) The zoom of the marked low temperature region of the specific heat figure in (a). We note that clearly	the} $p$-wave model fits the specific heat data much better throughout the temperature window than the other two models. (c) The low temperature specific heat shows a clear $\sim T^3$ temperature dependence characteristic of the presence of bulk point nodes as in the $p$-wave model. Here, the values of the constant parameters used in the fitting are: $a = 15.49$ and $b = 6.51$.}
	\label{fig:specific_heat}
\end{figure}
%%%%%%%%%%%%%%%%%%%%%%%%%%%%%%%%%%%%%%%%%%%%%%%%%%%%%%%%%%%%%%%%%%%%%%%
%%%%%%%%%%%%%%%%%%%%%%%%%%%%%%%%%%%%%%%%%%%%%%%%%%%%%%%%%%%%%%%%%%%%%%%

We have fitted the experimental data of $C_{el}$ for PdTe by the form given in \eqn{eqn:cv_theory} assuming a spherical Fermi surface, for three different types of models \ttd{\sout{as shown in the \fig{fig:specific_heat}(a) }}corresponding to the superconducting gap symmetries: (i) \textit{s}-wave with $g(\bk) = 1$ \ttds{having a full gap}, (ii) \textit{p}-wave with $g(\bk) = \sin(\theta)$ having only point nodes, and (iii) \textit{d}-wave with $g(\bk) = \sin(2\theta)$ having a line node at the equator ($\theta = \pi/2$) that dominates the low temperature behaviour~\cite{Biswas2021}. {We \ttds{first} make sure that the jump at ${\rm T}_{\rm c}$ in the electronic specific heat is \ttds{accurately} reproduced from all the fitted models}. \ttds{The fittings of $C_{el}/T$ vs. T, which enables us to look into the low temperature region carefully, are shown in \fig{fig:specific_heat}(a)}. We note from this \fig{fig:specific_heat}(a) that \ttds{the specific heat behaves as a power-law expected for a nodal order parameter instead of being exponentially suppressed expected for a fully gapped order parameter as a function of temperature in the low temperature region $0 \leq T/ {\rm T}_{\rm c} \lesssim 1/3$. As a result, the $\textit{s}$-wave does not give good fit to the data}. The $\textit{d}$-wave model also gives a poor fitting while a reasonably good fit to the experimental electronic specific heat data throughout the temperature range was obtained for the $p$-wave superconducting gap function. {To exemplify the low temperature behavior of the specific heat we have also shown a zoomed version of \fig{fig:specific_heat}(a) in \fig{fig:specific_heat}(b)}. The values of the fitting parameter $\frac{\Delta_0}{k_B {\rm T}_{\rm c}}$ for \ttd{\sout{fitting with }}the $s$-, $p$- and $d$-wave models shown in \fig{fig:specific_heat} are $\sim 1.83$, $\sim 2.29$ and $\sim 2.56$ respectively. Hence, we conclude that the electronic specific heat of PdTe is well described by a weak-coupling $\textit{p}$-wave generalized BCS-type order parameter with $\frac{\Delta_0}{k_B {\rm T}_{\rm c}} \sim 2.29$ which is close to the conventional weak-coupling BCS value $\sim 1.76$. This conclusion is further validated by \ttds{explicitly fitting} the low temperature behaviour of the specific heat \ttds{with temperature power laws expected for the nodal order parameters} as shown in the \fig{fig:specific_heat}(c). We note that there is a clear $\sim T^3$ temperature dependence in the region $0 \leq T/ {\rm T}_{\rm c} \leq 1/3$ characteristic of the presence of bulk point nodes in the superconducting order parameter of PdTe. This C$_{\rm el}$ $\propto$  T$^{3}$ dependence at low temperatures for PdTe is quite remarkable and conforms with the evidence of the bulk point nodes in the superconducting order parameter reported based on ARPES measurements~\cite{yang2023coexistence}. 

\ttds{PdTe is a multiband superconductor with the point group $D_{6h}$ which has both non-degenerate and degenerate irreducible representations. As a result, several symmetry allowed single- and multi-component superconducting order parameters are allowed in PdTe and in general it is difficult to clearly identify which superconducting instability channel is realized in PdTe. However,} mean-field calculations using a Slater-Koster type tight-binding model for PdTe, considering only the $p_x$ and $p_y$ orbitals of Te and the effect of spin-orbit coupling such that the Type-I Dirac crossing observed in ARPES along the $\Gamma-A$ direction is qualitatively captured, also suggest that a superconducting gap function with a bulk point node is indeed possible in PdTe \ttds{even in the most symmetric superconducting instability channel}~\cite{yang2023coexistence} . 

The jump in the electronic specific heat ($\Delta C$) at T$_{\rm c}$ in zero field ($H=0$) for PdTe shown in  the \fig{fig:specific_heat}(a) is  $\Delta C/(\gamma_{N} T_{\rm c}) \approx 1.49$ which is also similar to the weak coupling BCS value $\sim 1.43$~\cite{mcmillan1968transition}. Our finding is consistent with previous studies $\Delta C/(\gamma_{N} T_{\rm c}) \approx 1.33$~\cite{tiwari2015pdte} and $\approx 1.67$~\cite{karki2012pdte} but a little less compared to the value $\approx 2.1$ reported in the Ref.~\cite{chapai2023evidence}. The value of $\Delta C/(\gamma_{N} T_{\rm c})$ can, however, depend on the sample quality and the superconducting volume fraction in the samples. Moreover, some superconductors such as  many of the cuprate and Fe-chalcogenide superconductors do not exhibit jump in the specific heat as well~\cite{DingPRB2008,TropeanoPRB2008,SCHNELLE1993456}. We note that the electronic specific heat data of PdTe was reported to be fitted with an unconventional two-gap s-wave type model in the Ref.\cite{chapai2023evidence} where the corresponding samples had atleast $\sim 16\%$ non-superconducting volume fraction. \ttds{We also note that another recent work on PdTe reported multigap behaviour from thermal conductivity data~\cite{Zhao2024}. However, the thermal conductivity data may not be able to properly determine the nodal superconducting behaviour due to the corresponding small size and irregular shapes of PdTe single crystals used~\cite{Zhao2024}.}

\section{Summary and Conclusion}
Nodal superconducting materials~\cite{Gannon2015,Bhattacharyya2019,Shang2020,Biswas2021} represent a rare and intriguing class of unconventional superconductors, captivating the interest of researchers seeking to unravel their nature and pairing mechanisms. Based on recent angle-resolved photo-emission spectroscopy (ARPES) measurements~\cite{yang2023coexistence} it was proposed that PdTe, a 3D Dirac semimetal in its normal state, also exhibits point nodal superconductivity \ttds{in the bulk}. To verify the unconventional nature of superconducting order parameter in PdTe, we have conducted extensive field- and temperature-dependent specific heat measurements in the mixed state of PdTe. 

In our field dependent specific heat data, we have observed a power-law field dependence in the specific heat that deviates from the usual linear behavior expected for an isotropic fully gapped \ttd{\sout{s-wave type }}order parameter~\cite{Zehetmayer2013}. Our zero-field specific heat data offers additional insights, remarkably revealing a clear $\sim T^3$ temperature dependence in the low-temperature regime ($T <  {\rm T}_{\rm c}/3$) that strongly suggests the presence of bulk point nodes in the superconducting order parameter of PdTe. Notably, a weak-coupling BCS-type superconducting order parameter with \ttd{\sout{p-wave symmetry and }}point nodes yields an excellent fit to the zero-field specific heat data throughout the temperature range. Thus, our bulk specific heat measurements in the superconducting state of the 3D Dirac semimetal PdTe point towards its rare identification as a point nodal superconductor \ttd{\sout{and emphasizes PdTe to be a potential topological superconductor }}as proposed in the Ref.~\cite{yang2023coexistence}. 

{PdTe stands out as a unique 3D Dirac semimetal material~\cite{yang2023coexistence} because it: i) features Dirac points near the Fermi level, ii) has an accessible transition temperature ($T_c$), iii) is cleavable to measure superconducting Fermi arc states, and iv) has unconventional point nodal superconducting ground state.} Further confirmation of the point nodal superconducting order parameter in PdTe will require the use of complementary experimental techniques, such as solid-state nuclear magnetic resonance (NMR) to measure spin-susceptibility~\cite{Smith2006} and tunnel-diode oscillator (TDO) to measure the London penetration depth~\cite{giannetta2022london} {in combination with orientation-dependent measurements with good quality single crystals~\cite{Mirano2003, Vekhter1999}}.

\section{Acknowledgment}

CSY acknowledges SERB-DST (India) for the CRG grant (CRG/2021/002743). Advanced Material Research Center (AMRC), IIT Mandi is acknowledged for the experimental facility. SKG gratefully acknowledges financial support from SERB, Government of India via the Startup Research Grant: SRG/2023/000934, as well as from IIT Kanpur through the Initiation Grant (IITK/PHY/2022116). The authors also acknowledge Dibyendu Samanta for discussions.\\
 
\bibliography{PT}

\end{document}